# Efficient Post-processing of Diffusion Tensor Cardiac Magnetic Imaging Using Texture-conserving Deformable Registration

Fanwen Wang[a,b,c], Pedro F.Ferreira[b,c], Yinzhe Wu [a,b,c], Camila Munoz[b,c], Ke Wen[b,c], Yaqing Luo[b,c], Jiahao Huang[a,b,c], Dudley J.Pennell [b,c], Andrew D. Scott [b,c], Sonia Nielles-Vallespin [b,c] and Guang Yang*[a,b,c,d]

[a]Bioengineering Department and Imperial-X, Imperial College London, London W12 7SL, UK;
[b]National Heart and Lung Institute, Imperial College London, London SW7 2AZ, UK;
[c]Cardiovascular Magnetic Resonance Unit, Royal Brompton Hospital, Guy's and St Thomas' NHS Foundation Trust, London SW3 6NP, UK;
[d]School of Biomedical Engineering & Imaging Sciences, King's College London, London WC2R 2LS, UK;

## ABSTRACT

Diffusion tensor cardiac magnetic resonance (DT-CMR) is a method capable of providing non-invasive measurements of myocardial microstructure. Image registration is essential to correct image shifts due to intra and inter breath-hold motion and imperfect cardiac triggering. Registration is challenging in DT-CMR due to the low signal-to-noise and various contrasts induced by the diffusion encoding in the myocardium and surrounding organs. Traditional deformable registration corrects through-plane motion but at the risk of destroying the texture information while rigid registration inefficiently discards frames with local deformation. In this study, we explored the possibility of deep learning-based deformable registration on DT-CMR. Based on the noise suppression using low-rank features and diffusion encoding suppression using variational auto encoder-decoder, a B-spline based registration network extracted the displacement fields and maintained the texture features of DT-CMR. In this way, our method improved the efficiency of frame utilization, manual cropping, and computational speed.

## 1. DESCRIPTION OF PURPOSE

Diffusion tensor cardiac magnetic resonance (DT-CMR) is a non-invasive imaging modality that reveals the function of cardiomyocytes. After acquiring data with different diffusion encoding directions a tensor can be calculated based on the self-diffusion of water molecules in the myocardium. To acquire one slice, multiple repetitions of different directions need to be collected to increase the signal-to-noise ratio. Despite using cardiac triggering and breath-holding to reduce motion effects, inconsistencies in breath control and heart rate variability can still induce motion artifacts, especially during the lengthy acquisitions.

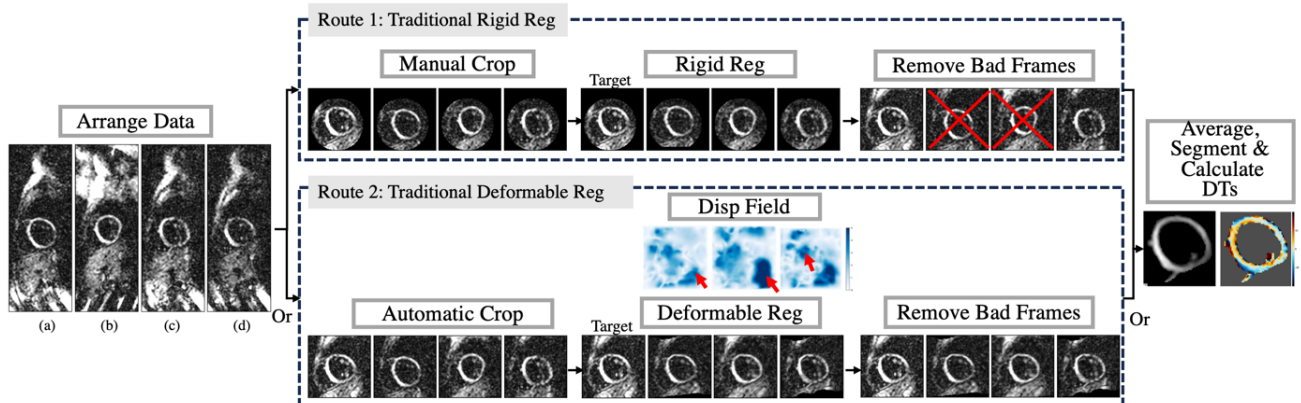

Figure 1 The post-processing pipeline of DT-CMR with details on registration. Reg denotes registration, Dis denotes the displacement field. The severe changes in the displacement field, pointed out by red arrows, indicate the destruction of texture information.

Image registration is a process that finds the mapping from one image to another and aligns them in one coordinate system. Registration in DT-CMR poses a significant challenge due to the complex respiratory and cardiac motion, the variability of heart rates, as well as the heterogeneity of signal intensities and contrasts present in both the myocardium and surrounding chest wall and stomach[1]. Any misalignment induces bias in the final parameter estimation. Most DT-CMR pipelines include manual cropping to exclude the signal from neighboring organs, rigid registration of the image frames, and manual removal of bad frames due to poor registration performance, low SNR and signal loss (Fig 1, Route 1 panel). Although the imperfect cardiac triggering and breath-hold can induce the non-rigid motion, deformable registration methods, which rely on intensity-based or information-theoretic metrics, are often incompatible in DT-CMR. This difficulty is predominantly due to varying contrasts and intrinsically low SNR induced by diffusion encoding in the myocardium. The red arrows in the displacement field show the undesirable warping of the texture information (Fig 1, Route 2 panel). The mapping of intensity within the myocardium performed by most registration methods makes the frames anatomically alike but destroys the embedded texture information. However, simple rigid registration cannot correct the deformable motion leading to a lower efficiency in frame utilization.

In this study, we explored the possibility of using deformable registration in DT-CMR post-processing to improve efficiency. According to the underlying physics, DT-CMR can be modeled as a combination of anatomical tissue property, diffusion encoding information, and acquisition noise. By suppressing the noise using PCA-based denoising method and the diverse contrasts induced by diffusion encoding using a beta variational autoencoder, we disentangled the anatomy and contrast of DT-CMR. Then a diffeomorphic B-spline deep learning network on the different anatomies but similar contrast images were used to derive the smooth displacement field. The displacement fields were applied on the original noisy frames with texture embedded for a better tensor fitting. The proposed method enhances efficiency through three key improvements: it optimizes frame utilization, obviates the necessity for manual cropping, and significantly boosts computational speed.

## 2. METHOD

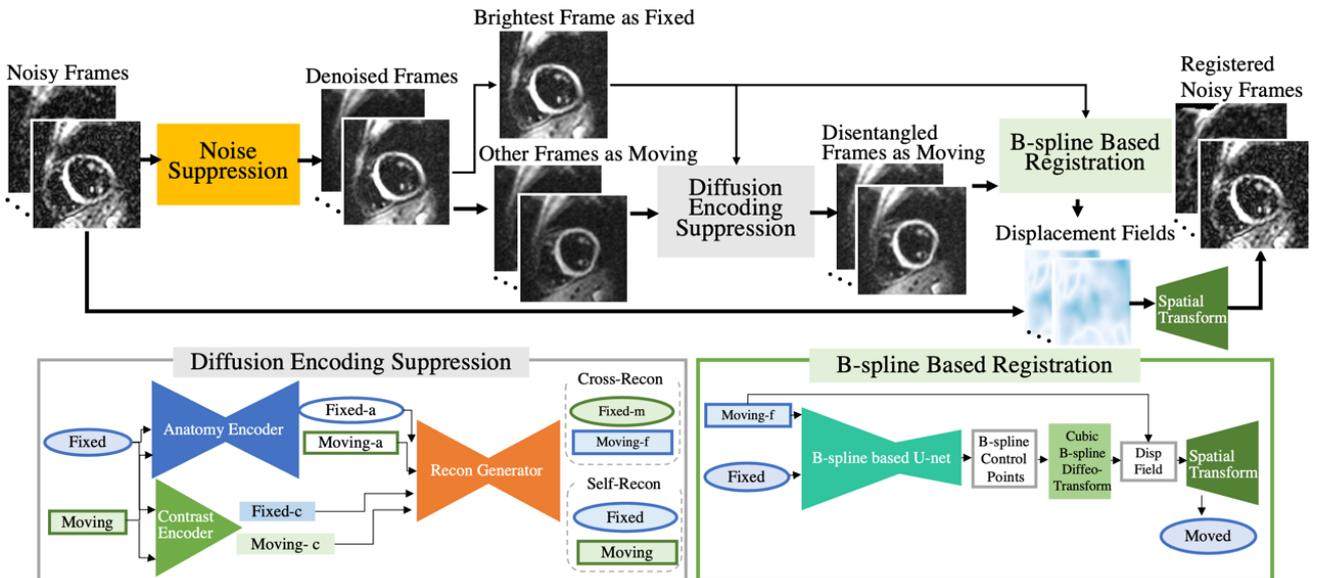

Figure 2 The architecture of the texture-conserving network. -c and -a denote the disentangled contrast and anatomical information. -f and -m denote the cross-reconstruction using the contrast from fixed or moving frames respectively.

**2.1 Data Acquisition and Preparation**

20 healthy volunteers were scanned at 3T and 1.5T using SIEMENS MRI scanners with a Stimulated Echo Acquisition Mode (STEAM) based Echo Planar Imaging (EPI) sequence with a spatial resolution of 2.8×2.8×8 mm$^3$. *b0* data and 6 encoding directions at b = 150 (b150) and 600 s/mm² (b600) with 10.0 [10.0 12.0] in short-axis midventricular slice at both end-systole (ES) and end-diastole (ED) were collected. A total of 20 (volunteers) × 2 (1.5 T/3T) × 2 (ES/ED) = 80 cases were included. Since every scan had different orientations and volunteer positions, we treated each case as independent. For a fair comparison, all the frames of 7 averages of b600 acquisition and 2 averages of b150 were used. The training, validation, and testing dataset included 65, 5, and 10 cases respectively. All the images were cropped to the same matrix size of 96×96 of the central part. The images were normalized using the max intensity of the brightest frame per case, and input to the diffusion encoding suppression and registration.

## 2.2 Rationale and Network Architecture

According to the underlying physics[2], DT-CMR can be modelled as a combination of anatomical tissue property $a$, diffusion encoding information $c$, and acquisition noise $n$.

$$d = nca \qquad (1)$$

**Noise Suppression:** We first developed a groupwise Principal Component (PC) analysis denoising[3] method to suppress noise and retain the diffusion encoding information (Fig 2). The PCs with 97% of the signal information were included with an autocorrelation function for genuine diffusion effects. The rejected PCs were set to zero and multiplied by corresponding PC scores to generate denoised frames.

$$\boldsymbol{d} = \boldsymbol{m} + \boldsymbol{T} * \boldsymbol{W^T} \qquad (2)$$

where $\boldsymbol{d}$ represents all the frames in one case, $\boldsymbol{m}$ represents the mean signal, $\boldsymbol{T}$ is the score vector and $\boldsymbol{W^T} = [W_1^T, ..., W_M^T]$ is a basis matrix with columns as PC vectors. M = 63 is the number of frames. The power of the first 1/3 PCs was calculated, and a second order polynomial was fitted to estimate the noise. The PCs with 97% of the signal information were included. An autocorrelation function was used to include the PCs with positive values which stood for genuine diffusion effects:

$$R(l) = \frac{1}{(M-l)Cov(\boldsymbol{d})} \sum_{i=1}^{M}(W_i - W)(W_{i+1} - W) \qquad (3)$$

$Cov(\boldsymbol{d})$ is the covariance of all frames. The PCs with positive autocorrelation with delays $l \leq 3$ were also incorporated as diffusion information. Then the rejected PCs were set to zero and multiplied by corresponding PC scores to generated denoised slices.

**Diffusion Encoding Suppression:** We implemented a deep learning based disentanglement network[4] for the denoised frames in a pairwise manner (Fig 2). The frame with the highest intensity was chosen as the fixed frame $d^F$ and all the other frames as moving $d^M$. Both included the contrast information of $c$ and anatomical information of $a$

$$d^F = c^F a^F, \ d^M = c^M a^M \qquad (4)$$

Variational encoders extracted anatomy information ($a^F$, $a^M$) and the mean and contrast information ($c^F$, $c^M$). Then we synthesized the frames using self-reconstruction and cross-reconstruction and took the cross-reconstructed image $d^{FM}$ with the initial anatomy $a^M$ but similar $c^F$ as the input of the registration network.

$$d^{FF} = c^F a^F, \ d^{MM} = c^M a^M, \ d^{FM} = c^F a^M, \ d^{MF} = c^M a^F \qquad (5)$$

Losses for self-reconstruction, cross-reconstruction, anatomy similarity loss and contrast bottleneck loss were utilized to supervise the disentanglement network:

$$L_{self-recon} = \|d^{FF} - d^F\|_1 + \|d^{MM} - d^M\|_1 \qquad (6)$$

$$L_{cross-recon} = \|d^{FM} - d^F\|_1 + \|d^{MF} - d^M\|_1 \qquad (7)$$

$$L_{anatomy} = 1 - \frac{<a^M, a^F>}{\|a^M\|_2 \cdot \|a^F\|_2} \qquad (8)$$

$$L_{contrast} = \| \|c^M\|_2 - c^F \|_1 \qquad (9)$$

$$L_{overall} = \lambda_1(L_{self-recon} + L_{cross-recon}) + \lambda_2 L_{perce} + \lambda_3 L_{anatomy} + \lambda_4 L_{contrast} \qquad (10)$$

**B-spline Based Registration:** To suppress undesirable drastic displacement fields, we adopted a diffeomorphic B-spline based registration network[5]. Based on a variant of U-net, we generated the control points of the image (Fig 2). With a differentiable mutual information metric, the registration network succeeded in generating a smooth displacement field for registration. The loss was defined as $L_{overall}$:

$$L_{NMI}(d^F, d^{FM} \circ \psi) = \frac{\sum_M H(d^F, d^{FM} \circ \psi) + \sum_{FM} H(d^F, d^{FM} \circ \psi)}{H(d^F, d^{FM} \circ \psi)} \quad (11)$$

$$\lambda_{Reg} = \sum_m \left\| \frac{d(\psi(t))}{dm} \right\|_2^2 \quad (12)$$

$$L_{overall}(d^F, d^{FM} \circ \psi) = L_{NMI}(d^F, d^{FM} \circ \psi) + \lambda L_{Reg} \quad (13)$$

where $L_{NMI}$ is a normalized mutual information and $\lambda_{Reg}$ is a L2 regularization on velocity field $\psi(t)$. $H(d^F, d^{FM} \circ \psi) = \sum_{M \cap FM} p(d^F, d^{FM} \circ \psi) \ln p(d^F, d^{FM} \circ \psi)$ denotes the joint entropy with joint histogram of $p(d^F, d^{FM} \circ \psi)$. $M \cap FM$ denotes the overlapping area and $m$ denotes the spatial dimension of *(x, y)*. A Parzen window[5] was applied on the joint histogram to make the loss differentiable. After extracting the desired displacement field $\psi$, we set the brightest frame as fixed and registered all the other frames accordingly.

**Final implementation:** After we have removed the noise and contrast from the diffusion weighted frames which then allowed registration algorithms not to be confused by these additional factors. The deformable fields were applied to the original data for a better tensor fitting.

## 2.3 Experiments Details

The proposed texture-conserving (Texture-DL) network was compared with traditional rigid (Rigid) and B-spline based deformable registration (Deformable) using Elastix[6]. Ablation studies used DL registration on noisy frames (Noisy-DL), and DL on denoised frames (Denoised-DL).

Rigid registration used Matlab function dftregistration[7], which took a cross correlation in Fourier space and discreate Fourier transform to allow upsampling of the images with less computation time.

Traditional deformable registration was implemented by a 2-level resolution B-spline registration using Python ITK-Elastix[6] package, with control point spacing = 16, B-spline interpolation order = 1, loss function of advanced Mattes mutual information and optimisation method of stochastic gradient descend.

The disentangled network and the diffeomorphic network were implemented using Pytorch Library v1.12.1 on RTX3090. For the disentangled network, we set epoch = 100 with step per epoch = 20, $\lambda_1 = 1$, $\lambda_2 = 0.03$, $\lambda_3 = 0.02$, $\lambda_4 = 10^{-8}$. For the DL-based registration network, we set $\lambda = 0.1$, control point spacing = 4 and epoch = 100. Both networks chose the model with epoch of the best performance on the validation set.

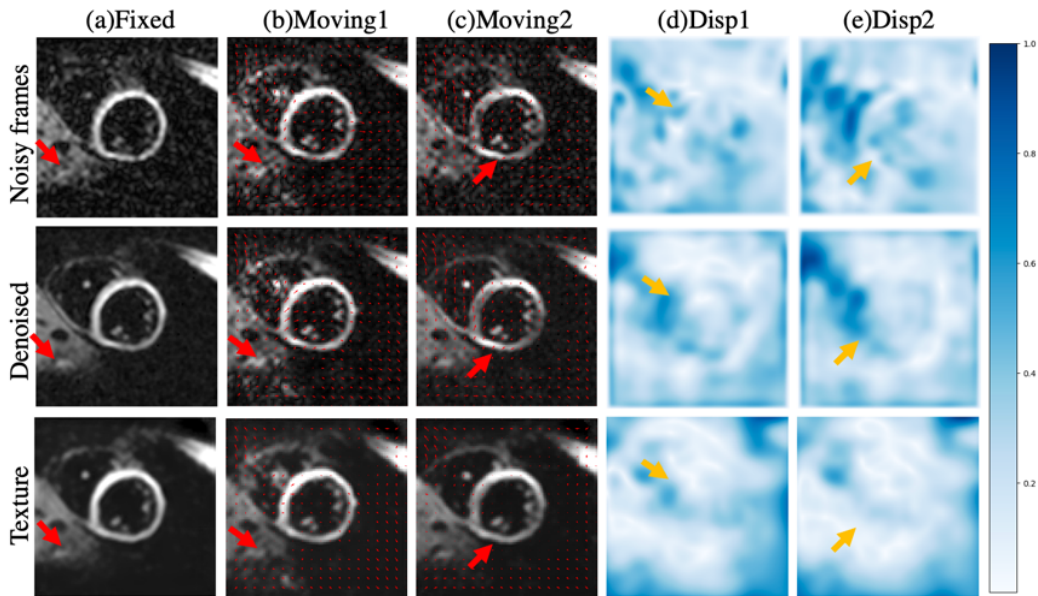

Figure 3 The displacement fields in ablation studies. Red arrows point out the diverse contrast in the surrounding organs and myocardium. Yellow arrows show the corresponding drastic deformation.

## 2.4 Evaluation Criteria

**Number of negative eigenvalues:** Three orthogonal eigenvectors with eigenvalues were extracted from the symmetric and positive-definite matrix diffusion tensor. Three eigenvalues give the diffusivity along the direction of each eigenvector and should be positive. Hence, the number of negative eigenvalues revealed number of invalid pixels with the effects of motion, noise or mis-registration artifacts[8, 9].

**Image stack visualization:** All acquisitions of the same slice were stacked along a third dimension. The central line from horizontal and vertical perspectives were chosen for qualitative visualization.

**Helix angle (HA) gradient line profile:** Without ground truth of displacement field, we followed the line profile gradient of helix angle[10] on healthy volunteers to evaluate the performance of registration. Under the assumption that the change from endocardium to epicardium be linear, we did regressions for each line along the radius direction of myocardium[6]. Line profiles with negative slopes and fitting $R^2$ larger than 0.3 were included. $R^2$ and root-mean-square error (RMSE) of the linear regression were compared.

## 3. RESULTS

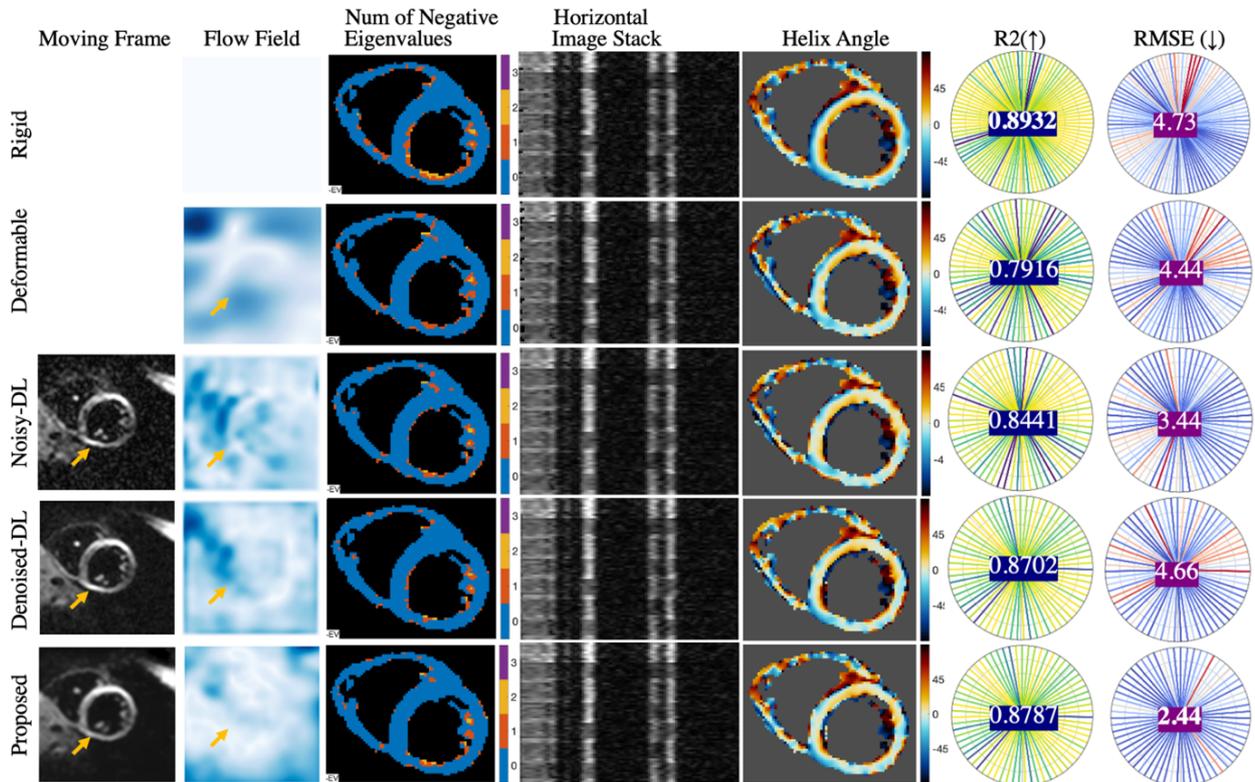

Figure 4 The visualization of the diffusion tensor parameters with R2 and RMSE of the fitting of helix angle (HA) line profiles. HA is measured in degrees.

The output of images using denoising and disentanglement with corresponding displacement fields were shown (Fig 3). Compared with the original frames, denoised ones were less noisy with smoother displacement fields. The contrast within the myocardium in the moving frames (c) differed from the fixed frame (a). Moreover, the contrast in the bottom right of the myocardium, pointed by a red arrow, was suppressed using the disentanglement with smaller deformation in the displacement field. Compared with the registration networks applied on the noisy frames or denoised frames, the texture-conserving network showed less undesirable deformation within the myocardium, pointed by yellow arrows.

The tensor parameter estimations of different methods were evaluated (Fig 4 and Table 1). The proposed method had the lowest RMSE, and highest $R^2$ with the least percentage of negative eigenvalues.

Table 1. Comparison and Ablation methods. Negative denotes the percentage of pixels with negative eigenvalues. R-squared and RMSE are shown with mean and standard deviation. Percentages are shown using median and 25% interquartile and 75% of interquartile.

| Methods | R-squared (↑) | RMSE(↓) | Negative (↓) |
|---|---|---|---|
| Rigid | 0.876±0.155 | 6.509±5.265 | 4.4 [2.7, 7.4] |
| Deformable | 0.850±0.164 | 6.557±5.006 | 5.0 [1.6, 7.2] |
| Noisy-DL | 0.844±0.164 | 6.900±5.494 | 5.5 [2.6, 8.2] |
| Denoised-DL | 0.857±0.161 | 6.769±5.645 | 4.3 [2.9, 7.2] |
| **Proposed** | **0.881±0.139** | **5.716±4.560** | **2.7 [1.2, 4.6]** |

## 4. CONCLUSIONS

Poor cardiac triggering and breath-hold induce deformable motion in DT-CMR. However, its low SNR and diverse contrasts make most registration with intensity-based or information-theoretic metrics incompatible. In this study, we removed the noise and contrast from the diffusion weighted images using the PCA-denoised method to suppress noise and disentangling network respectively. Then a diffeomorphic B-spline registration network with differentiable mutual information were applied on the disentangled DT-CMR with similar contrast but different motion-induced anatomies. The displacement fields were then extracted and applied on the original diffusion weighted images for better tensor calculations. Results showed that compared with traditional rigid and deformable registration methods, the proposed methods achieved the highest $R^2$, least RMSE in linear fits to transmural HA, and lowest percentage of number of negative eigenvalues on healthy volunteers.

# SUPPLEMENTARY FILES

We further collected data using a navigator-based STEAM-EPI sequence to test the performance of the proposed method on data with large motion.

Dataset 2 consists of two short-axis slices (apical and basal) at end-systole (ES) from 20 healthy volunteers. These images were acquired using a 3T MRI scanner with STEAM-EPI. Each slice includes 12 repetitions of b0, 2 repetitions of b150, and 10 repetitions of b600 images. The image resolution is 2.8 × 2.8 × 8.0 mm³. One-dimensional navigators with a 4mm acceptance window were placed on the diaphragm to reject frames with significant through-plane motion. This results in a total of 40 cases (20 volunteers × 2 slices each).

We conducted experiments using the existing dataset and dataset 2. The combined dataset now includes 120 cases, with 40 low-quality cases (determined by poor segmentation performance) used as the test dataset. Different from just collecting 63 frames for each case, we utilized all the frames to meet the clinical requirement.

The results were as follows:

Table S1. Results on two datasets. Negative denotes the percentage of pixels with negative eigenvalues. R-squared and RMSE are shown with mean and standard deviation. Percentages are shown using median, 25% and 75%.

| Methods | R-squared (↑) | RMSE(↓) | Negative (↓) |
|---------|---------------|---------|--------------|
| Proposed | 0.895±0.054 | 6.865±3.481 | 1.7 [0.6, 3.1] |